# Charged Particles Multiplicity and Scaling Violation of Fragmentation Functions in Electron-Positron Annihilation

Tooraj Ghaffary[1]
1- Department of Science, Shiraz Branch, Islamic Azad University, Shiraz, Iran
E.mail: tooraj.gh@gmail.com

**Abstract** In electron – positron annihilations, changing quarks into the final hadron states is being described by fragmentation function. Based on the scaling hypothesis, this function will be independent from the center of mass energy. Gluon radiation violates the scaling feature. Transverse momentum distribution plays an important role in scaling violation of fragmentation functions. In this article, by the use of data resulted from the annihilation process of electron – positron in AMY detector at 60 GeV center of mass energy, first, charged particles multiplicity distribution will be obtained and it will be fitted with the KNO scaling. Furthermore, momentum spectra of charged particles and momentum distribution respect to the jet axis will be obtained. Then, the results will be compared regarding the different models of QCD; as well, the distribution of fragmentation functions and scaling violations will be studied. It is being expected that the scaling violations of the fragmentation functions of gluon jet are stronger to the quark one. One of the reasons for such case is that splitting function of jet is larger than splitting function of gluon.

## 1- Introduction

Hadron production in high energy interactions can be described by the parton cascade [1] [the propagation of gluons and their separation into partons], and it is not possible to describe the formation of hadrons as perturbation description. Gluon radiation which is a prominent process in the parton cascade proportional with the color coefficient of radiated gluon coupling. This coefficient equals $C_A = 3$ when it radiates gluon, but when it radiates quarks, it will be $C_F = \frac{4}{3}$ [2]. As a result, the multiplicity of soft gluons from a gluons source is about 9/3 times bigger than the multiplicity of quark source. Inequality of $C_A$ and $C_F$ plays an important role in the explanation of the observed differences between the gluon and quark jets. Compared with quarks' jets, it is being observed that gluon jets have the high width, more multiplicity, soft fragmentation functions and strong scaling violations of fragmentation functions [3]. The fragmentation function "$D_a^h(x, Q^2)$" shows this possibility as parton "a" which is being produced in the short distances might be in the range of $\frac{1}{Q}$ and fragment into hadron "h" and also it has "x" fraction from a momentum of parton "a" [4-7]. In LEP experiments, momentum fraction is $x_E = \frac{E_h}{E_{jet}}$ in which $E_h$ shows the hadron energy of "h" and $E_{jet}$ refers to a jet energy which $E_h$ belongs to it [3]. Relative softness of fragmentation function of gluon jet in the area of small $x_E$ is being expressed by the multiplicity of radiated soft gluon; but in the other area with high value of $x_E$, it is being described by this fact as gluons cannot exist as a valence parton inside a produced hadron. The strong scaling violation of fragmentation functions of gluon's jet is resulted from this fact as the dependency of this scale (means the fragmentation functions of gluon jet by the separation function of $P_{g \to gg} \sim C_A$ ) is prominent; however, the dependency of fragmentation functions of quark jet by the separation function of $P_{q \to qg} \sim C_F$ is dominant. Given that, the momentum distribution of charged particles has a key role in the scaling violation of fragmentation functions, so in this article, first charged particles distribution will be obtained and will be fitted with the KNO scaling [11] in order to determine if it is consistent with other data from other energies. Then, momentum distribution respect to the jet axis will be obtained and at the end, distribution of fragmentation functions will be considered. Moreover, by using the data resulted from the AMY detector at 60 GeV center of mass energy, and comparing them with data obtained in other energies, Scaling violations of fragmentation functions will be explained.

## 2- Multiplicity of charged particles

Multiplicity distribution of charged particles is shown in figure 1. Figure 2 shows the probability distribution of charged particles which is indicated the cross section.

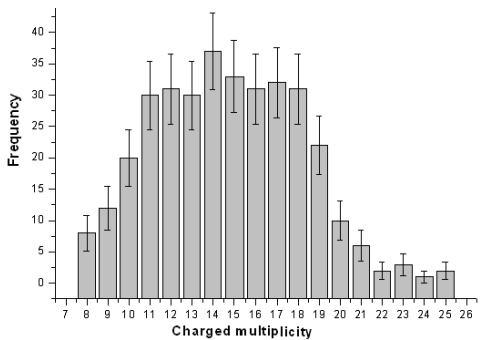 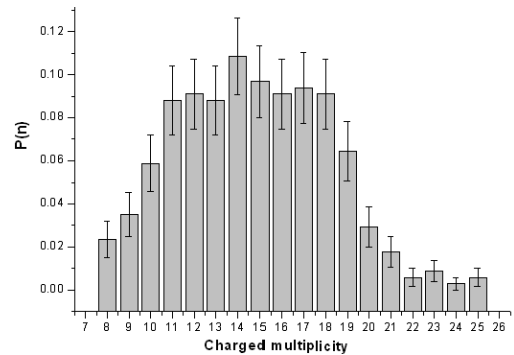

Figure 1. Frequency of charged particles multiplicity

Figure 2. Probability distribution of charged particles multiplicity

Average value of multiplicity of charged particles for AMY data is 14.68±3.83. The average value of multiplicity of charged particles in AMY data is consistent with the other data in the other energies and by increasing the energy, this value will be increased [9-12].

**3- KNO scaling**

An interesting description for the distribution of multiplicities in the definite energies was introduced by Koba, Nielsen and Olesen (KNO) which was derived from Feynman scaling [13-14]. The average of charged particles multiplicity, $\langle n_{ch} \rangle$, and its probability, $P(n_{ch})$, for the multiplicity of charged particles, "$n_{ch}$", in a function is defined as follows:

$$\Psi\left(\frac{n_{ch}}{\langle n_{ch} \rangle}\right) = P(n_{ch})\langle n_{ch} \rangle \tag{1}$$

For AMY data, this distribution can be drawn. For this reason, its figure will be as $\frac{n_{ch}}{\langle n_{ch} \rangle}$ based on $P(n_{ch})\langle n_{ch} \rangle$ which was shown in figure 3.

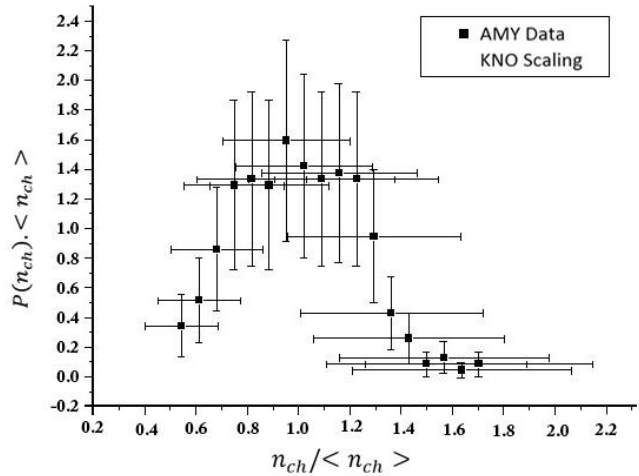

Figure 3. Measured charged particles multiplicity based on KNO scaling

Figure 3 has an acceptable consistency with the figure of other energies [15]. The scale function of KNO is being defined as follows [8]:

$$\Psi(z) = \frac{k^k}{\Gamma(k)} z^{k-1} e^{-kz} \tag{2}$$

$z = \frac{n}{\langle n \rangle}$ and $\Gamma(k)$ is Gamma function. AMY data was fitted with KNO function and its result was shown in figure 4.

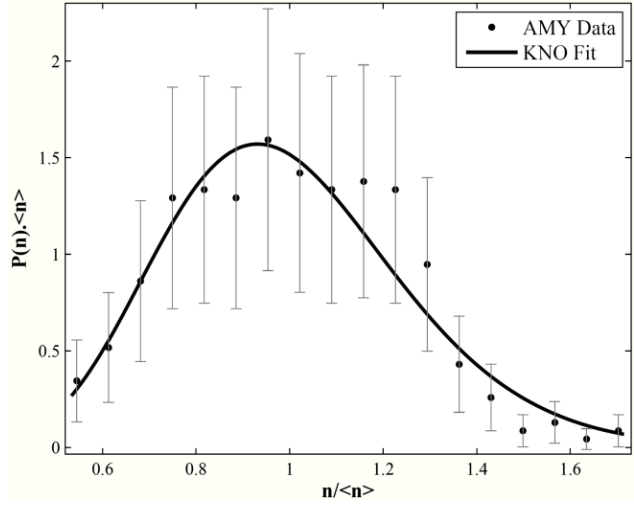

Figure 4. Charged multiplicity distribution due to AMY data based on KNO scaling

For $\frac{\chi^2}{N_{DF}} = 0.799$ , "K" equals to 14.62±2.98. Based on the existing errors in the tests, it can be shown that the value of K for AMY data is consistent with the value of the other tests. The magnitude of "K" has an interval between 11.64 to 17.60 [16].

**4- Momentum spectra of charged particles**

One of the features which is so effective in the interpretation of scaling violation is the effect of transverse momentum or $P_T$ on the fragmentation functions. For this reason, first, momentum spectra of charged particles to the jet axis for AMY data will be studied. Jets, in the annihilation process of electron – positron will be defined by the use of jet finder algorithm which is a mathematical meaning for dividing a phenomenon into the other parts which depend on the distinctive quarks and gluons. The most common algorithms are DURHAM [17] and JADE [18] jet finders. In this paper, JADE algorithm is used for finding jets. Using sphericity, the eigenvalues and eigenvectors of momentum tensor has been calculated [19]. A plane which is formed by the two eigenvectors corresponding to two bigger eigenvalue of momentum tensor,$(\hat{n}_1 - \hat{n}_2)$, is called as event plane. The average of transvers momentum in this plane, $\langle P_{Tin} \rangle$, is presented as follows:

$$\langle P_{Tin} \rangle = \frac{1}{N_{ch}} \sum_i \vec{P}_i \cdot \hat{n}_2 \qquad (3)$$

The summation is over on the charged particles. The average of transverse momentum in the direction perpendicular to the event plane,$\langle P_{Tout} \rangle$, was presented as the following equation:

$$\langle P_{Tout} \rangle = \frac{1}{N_{ch}} \sum_i \vec{P}_i \cdot \hat{n}_1 \qquad (4)$$

The summation is over on the charged particles. So, the average of total transverse momentum will be as follows:

$$\langle P_T \rangle = \sqrt{\langle P_{Tin}\rangle^2 + \langle P_{Tout}\rangle^2} \tag{5}$$

In figure 5, the cross section of $\frac{1}{\sigma_{tot}}\frac{d\sigma}{dp}$ related to the charged particles from AMY data in the energy of 60 GeV along with results from the lower energies in the various experiments are shown.

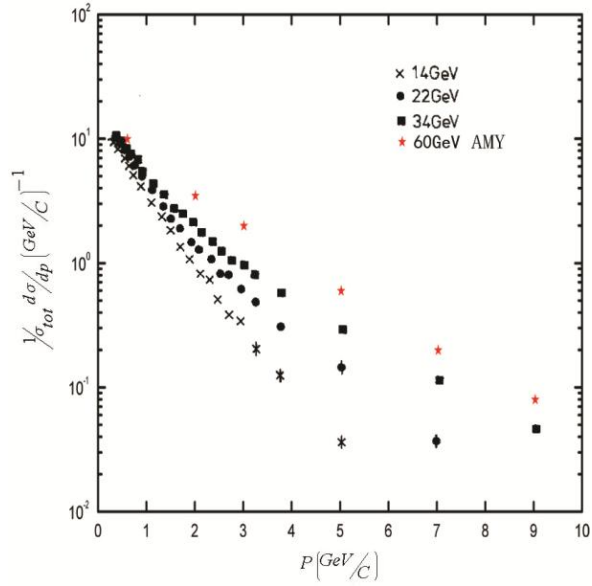

Figure 5. Cross section distribution related to the charged particles at different energies [20]

Based on figure 5, it can be seen that the cross section of particles' production for AMY data as well as the other data for p>0.2 will be decreased by increasing the momentum. So, by increasing energy, the widest curve will be shown by the momentum distribution. By decreasing the angle between two jets, the similarity of the event of three jets will be increased to the other event with two jets. One simple way for observing the gradual transfer [from the event of three jets into the event of two jets] is that the average distribution of transverse momentum. The square average of transverse momentum, in the event plane and perpendicular to this plane is presented as follows:

$$\langle P_{Tin}{}^2 \rangle = \frac{1}{N_{ch}} \sum_i \left(\vec{P}_i \cdot \hat{n}_2\right)^2 = Q_2 \langle P^2 \rangle \tag{6}$$

$$\langle P_{Tout}{}^2 \rangle = \frac{1}{N_{ch}} \sum_i \left(\vec{P}_i \cdot \hat{n}_1\right)^2 = Q_1 \langle P^2 \rangle \tag{7}$$

In figure 6, the square average of transverse momentum in the event plane, $\langle P^2_{Tin}\rangle$, and the square average of transverse momentum perpendicular to the event plane, $\langle P^2_{Tout}\rangle$, for AMY data along with the final results from different other energies were shown.

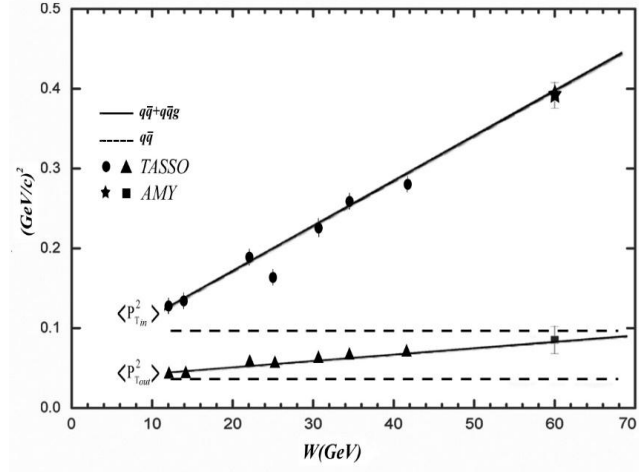

Figure 6. The square average of transverse momentum in the event plane, $\langle P^2_{Tin}\rangle$, and perpendicular to the event plane, $\langle P^2_{Tout}\rangle$, at different c.m. energies [21]

It is clear that $\langle P^2_{Tin}\rangle$ will increase much more strongly respect to $\langle P^2_{Tout}\rangle$. The results of QCD model regarding gluon radiation and without such radiation were shown in figure 6. It is being observed that AMY results with gluon radiation in QCD model are associated with more consistency. So, it can be concluded that there is the possibility of gluon radiation in high energies. Now, studying the transverse momentum distribution of charged particles to the jet axis is so important. Here, sphericity axis was selected as jet one. Distribution of transverse momentum, $(\frac{1}{\sigma_{tot}}\frac{d\sigma}{dP_T})$, and the distribution of squared transverse momentum, $(\frac{1}{\sigma_{tot}}\frac{d\sigma}{dP_T^2})$, for AMY data along with the final results in different energies were shown in figures 7 and 8. Therefore, for AMY data, $P_T$ and $P_T^2$ in comparison to the data resulted from the other experiments will show their widest distribution by increasing their energy. It is being expected that by increasing energy, multiplicity of particles will increase [22]. So, increasing the number of particles in $P_T \geq 0.5$, by increasing W, can be the reason of hard gluon radiation. In other words, this kind of radiation affects the flux of particles.

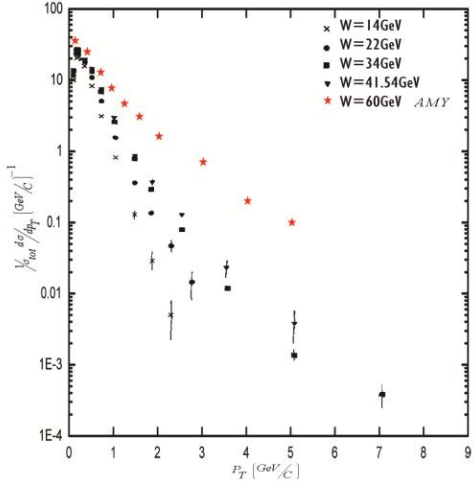 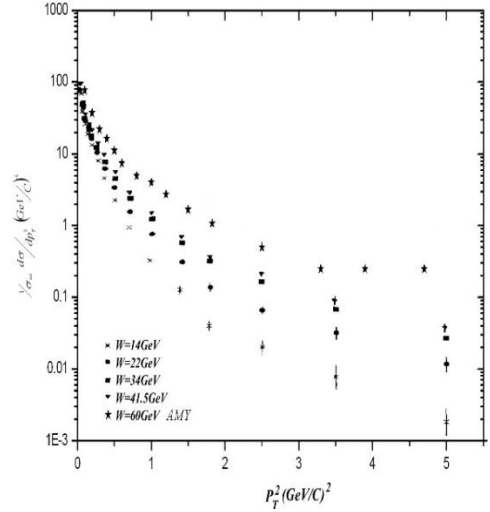

Figure 7. Distribution of transverse momentum, ($\frac{1}{\sigma_{tot}}\frac{d\sigma}{dP_T}$), at different energies [23]

Figure 8. Distribution of squared transverse momentum, ($\frac{1}{\sigma_{tot}}\frac{d\sigma}{dP_T^2}$), at different energies [21]

Now, the distribution of fragmentation functions and scaling violations will be studied.

## 5- Distribution of fragmentation functions and scaling violation

The fragmentation function is being defined as the whole number of charged particles, $N_{ch}$, in the bin related to $x_E$ and Q scale, which normalized to the number of jets "$N_{jet}(Q)$" [24-25]:

$$\frac{1}{N_{jet}(Q)}\frac{dN_{ch}(x_E, Q)}{dx_E} \tag{8}$$

In this equation, $x_E$ shows:

$$x_E = \frac{E_h}{E_{jet}} \tag{9}$$

A simple pattern of momentum distribution in terms of energy is important, when the energy of particles is being scaled as $x = 2E_h/W$. In such case, the possibility of using this variable was suggested by Feynman. This feature follows this rule:

$$\int D(x).x.dx = 2 \quad , \quad \int D(x).x°= n \tag{10}$$

In this case, "n" means the multiplicity of particles. In the model of quark and parton, cross section of $e^+e^- \to h + X$ is being defined by the following equation:

$$\frac{1}{\sigma_{tot}}\frac{d\sigma}{dx} = \frac{1}{\Sigma e_q^2}\sum e_q^2\left[D_q^h(x,s) + D_{\bar{q}}^h(x,s)\right] \tag{11}$$

In this domain, the summation is over on all quarks in which are being produced at $\sqrt{s}$ center of mass energy and $e_q$ shows the charge of quark. $D_q^h(x,s)$ defines a method in which the quarks will be changed into the final hadrons and it is called as the fragmentation function. As mentioned above, this feature shows the possibility of production hadron "h", with the energy scale of "x", from quark "q". In general, "D" depends on the kind of initial quark, hadron and the center of mass energy. Hadron production in the annihilation process of electron – positron can be expressed in the domain of structural functions of $F_1$ and $F_2$. Differential cross section of Yan, Levy and Drell, [26], is presented as:

$$\frac{d^2\sigma}{dx\, d\cos\theta} = \frac{3}{4}\sigma_\circ . x . \beta \left[ 2F_1(x,s) + \frac{x.\beta}{2} F_2(x,s).\sin^2\theta \right] \qquad (12)$$

$\sigma_\circ$ shows the cross section of QED in the $0^{th}$ order. So, $F_1$ and $F_2$ can be expressed in terms of the transverse and longitudinal structural function as:

$$F_T(x,s) = 2F_1(x,s) \qquad \text{(a-13)}$$

$$F_L(x,s) = 2F_1(x,s) + F_2(x,s) \qquad \text{(b-13)}$$

In which:

$$\frac{d^2\sigma}{dx\, d\cos\theta} = \frac{3}{4}\sigma_\circ . x \left[ F_T . (1+\cos^2\theta) + \frac{1}{2} F_L . \sin^2\theta \right] \qquad (14)$$

In comparison with the observed scaling behavior in the space-shaped distribution which was accounted as one of the accepted evidences for partons' presence, Yan, Levy and Drell assumed that $F_1$ and $F_2$ and also $F_L$ and $F_T$ were scaled as follows:

$$s \to \infty \quad \to \quad F(x,s) = F(x) \qquad (15)$$

In the framework of quark – parton model, photon is being coupled to a spin $\frac{1}{2}$ parton and:

$$F_L(x) = 0 \qquad \text{(a-16)}$$

$$xF_T = 3 \sum e_q^2 [D_q(x) + D_{\bar{q}}(x)] \qquad \text{(b-16)}$$

So, the scaling hypothesis shows that the fragmentation function of "D" must be independent from the energy of mass center. These formulas are not trusty in the framework of QCD. Gluon radiation leads into scaling violation and affects the longitudinal structural function. These changes are simple. Due to the radiation of a gluon, the energy of quark will decrease from $E_q$ to $E_{q'}$. Hadronic features of quark is being presented by a functional scale, $x' = E_h/E_{q'}$, instead of $x = E_h/E_q$ and for this reason, this quark will gain a momentum as its value in comparison to the evaluation of quark and parton will decrease (figure 9).

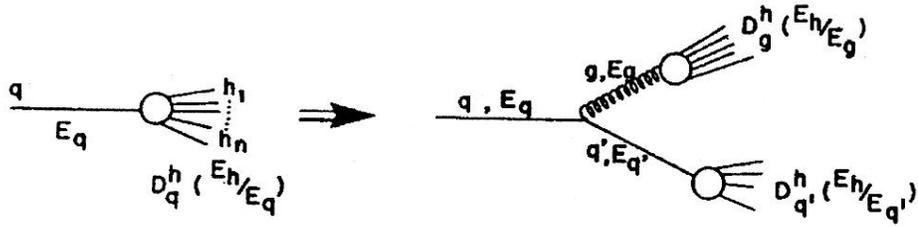

Figure 9. An overview of how a gluon radiation leads into scaling violation

So, the breakdown of QCD scale leads into the increase of particles in the lower Xs. But it is along with its decrease in the higher Xs. In addition, the effects of gluon change the features of angular momentum of the array of partons and it leads into a longitudinal element of $F_L$. In figures 10 and 11, the results related to scaling violations of fragmentation functions of quark and gluon jets along with the other results and AMY data were presented. These results are gained by considering the events of three jets.

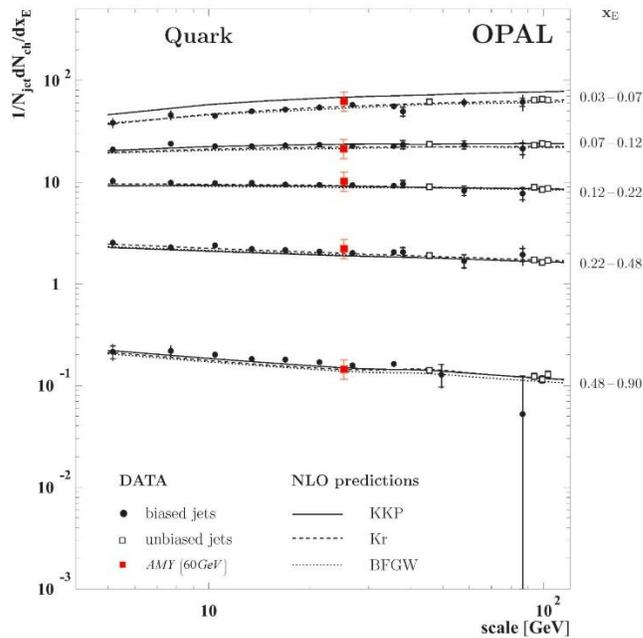

Figure 10. Scaling dependency of fragmentation functions of quark jets in $x_E$'s bins [25-28].

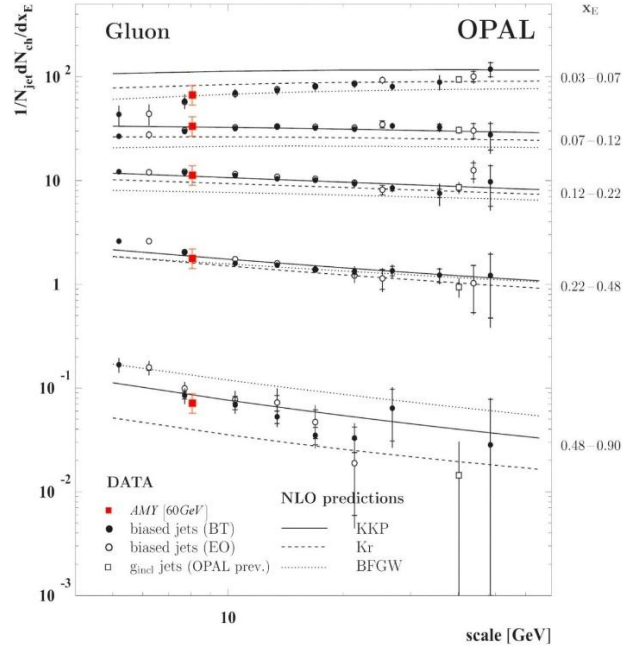

Figure 11. Scaling dependency of fragmentation functions of gluon jets in $x_E$'s bins [25-28].

For the fragmentation functions of quark jet (figure 10), all theoretical evaluations (NLO evaluations) are presented a good description for AMY data and other results except the smallest and largest variable of $x_E$. For the fragmentation functions of gluon jet, (figure 11), data explanation by NLO evaluations is not good for description. So, based on the above figures, scaling violation is observable as these violations for small $x_E$ are along with the positive slope and for large $x_E$ are along with the negative slope. In other words, it is being expected that stronger scaling violations will occur in the gluon jets to the quark ones. It is clear that the data in the small area of $x_E$ are more than KKP evaluations. So, in this area, the data is consistent with the evaluations of Kr and BFGW. For the large values of $x_E$, data is consistent with KKP evaluations. And the differences between these models will be decreased by increasing the related scales. In figures (12) and (13), the dependency of the fragmentation functions of quark and gluon jets to $x_E$ for AMY data was shown with the other results. In these two figures, AMY data in small $x_E$ is consistent with the theoretical evaluations and also with the other data from other energies. By increasing $x_E$, the cross section shows its descending trend as this trend is different for the fragmentation functions of gluon jets for various energies. So, by increasing energy, the dependency of cross section of these jets to energy will be increased. And for this reason, the scaling violation of fragmentation functions of gluon jets to the quark ones in the high energies is stronger.

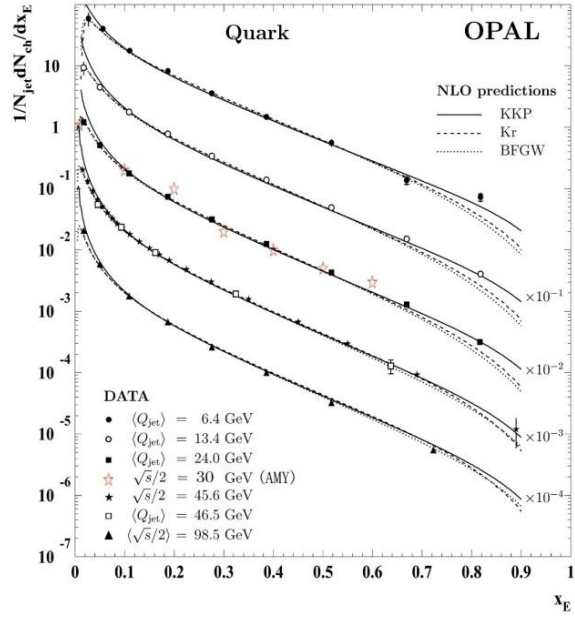

Figure 12. Dependency of the fragmentation functions of quark jets to $x_E$ in different scales at different c.m. energies [25-28].

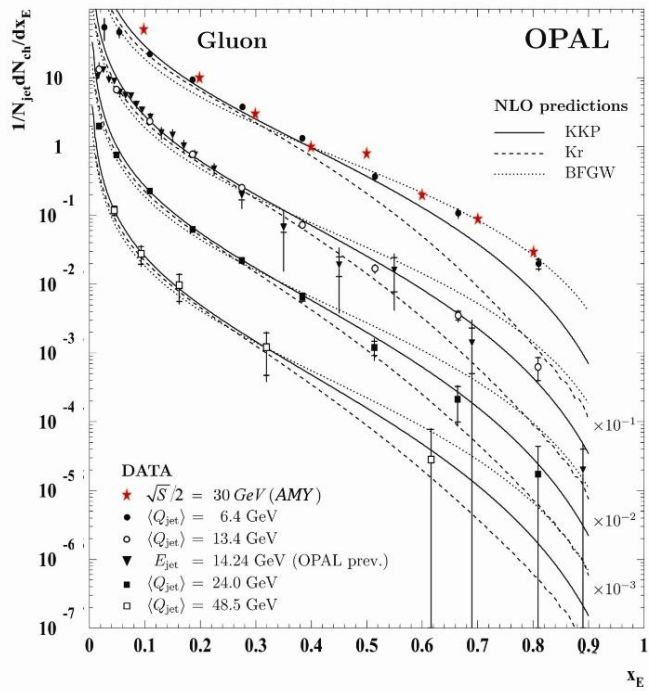

Figure 13. Dependency of the fragmentation functions of gluon jets to $x_E$ in different scales at different c.m. energies [25-28].

## 6- CONCLUSION

In this paper, by using resulted data from the annihilation process of electron – positron in AMY detector at 60 Gev center of mass energy. First, the momentum spectrum of charged particles to the jet axis will be studied as the sphericity axis was selected as the jet axis. Then, we notice the cross section of particles production will be decreased by increasing the momentum. In addition, by increasing such energy, this distribution becomes more width. In the wide range of energies, increase of average of momentum occurs in a liner trend. It must be noted that the square average of particles' momentum shows the considerable increase by increasing its energy. Its reason is that the possibility of gluon radiation in the high energies is so high. In other words, scaling violation of fragmentation functions occur in the higher energies with high possibility. It is being observed that in the large $P_T$s, by increasing the center of mass energy, "W", the number of particles will increase as the radiation of gluon can be a main reason for such case. In other words, radiation of gluon affects the flux of particles in this domain. And also, the distribution of fragmentation functions is being studied. As well, the violations of scaling are clear in these distributions as these violations for the fragmentation functions of gluon jets to the quark ones in the high energies is so high. At the end, it is determined that the reason of the scaling violations in the fragmentation functions by increasing energy is that the possibility of gluon radiation in the high energy is great. This result is consistent with the prediction of QCD.

**Acknowledgements**
The author would like to thank Islamic Azad University for fruitful collaborations. Financial support of this study has been made by Islamic Azad University-Shiraz branch.